\documentclass[a4paper, 12pt, openbib ]{article}
\usepackage{times}
\usepackage{pifont}
\usepackage{amsmath}
\usepackage{amssymb}
\usepackage{amsfonts}
\usepackage{graphicx}
\usepackage{floatflt}
\usepackage{geometry}
\usepackage{layout}
\usepackage{enumerate}
\usepackage{bm}
\usepackage{tikz}
\usetikzlibrary{patterns}
\def\tfract#1/#2{{\textstyle{\raise0.8pt\hbox{$\scriptstyle#1$}\over%
\hbox{\lower0.8pt\hbox{$\scriptstyle#2$}}}}}
\def\mezzo{\tfract 1/2 }
\def\imezzi{\tfract i / 2}
\def\terzo{\tfract 1/3}

\def\sesto{\tfract 1/6 }

\def\radi2k{\tfract 1/{\sqrt {2k}} }
\def\der{\partial }
\def\cvd{\vbox{\hrule \hbox to 9 pt {\vrule height 9 pt \hfil \vrule} \hrule}}
\def\downnormalfill{$\,\,\vrule depth4pt width0.4pt
\leaders\vrule depth 0pt height0.4pt\hfill\vrule depth4pt width0.4pt\,\,$}
\def\WT#1{\mathop{\vbox{\ialign{##\crcr\noalign{\kern3pt}
      \downnormalfill\crcr\noalign{\kern0.8pt\nointerlineskip}
      $\hfil\displaystyle{#1}\hfil$\crcr}}}\limits}

\def\be{\begin{equation}}
\def\ee{\end{equation}}
\def\bes{\begin{equation*}}
\def\ees{\end{equation*}}
\def\bea{\begin{eqnarray}}
\def\eea{\end{eqnarray}}
\def\beas{\begin{eqnarray*}}
\def\eeas{\end{eqnarray*}}
\def\ba{\begin{array}{rcl}}
\def\ea{\end{array}}
\def\der{\partial}
\numberwithin{equation}{section}
\def\go{\leavevmode \raise.3ex\hbox{$\scriptscriptstyle \langle\!\langle\!  $}%
~\ignorespaces}
\def\gf{\relax \ifhmode \unskip~\else \leavevmode \fi \raise.3ex\hbox{$\! \scriptscriptstyle\rangle\!\rangle\, $}}
\title{
{\Large  Schwinger-Dyson functional in Chern-Simons theory}
{\vskip 0.5 truecm}} 

\author{{\large  E.~Guadagnini}  \\  {\normalsize Dipartimento di Fisica {\it E. Fermi} dell'Universit\`a di Pisa,} \\ {\normalsize and INFN Sezione di Pisa.}  \\ {\normalsize Largo B. Pontecorvo  2, 56127 Pisa, Italy}}
\date{}

\begin{document}

\maketitle 

\vskip 0.7 truecm 

\begin{abstract}

In perturbative $SU(N)$ Chern-Simons gauge theory, it is shown that the Schwinger-Dyson equations assume a quite simplified form.  The generating functional of the correlation functions of the curvature is considered;   it is demonstrated that the  renormalized Schwinger-Dyson functional is  related with the generating functional of the correlation functions of the gauge connections by some kind of duality transformation.   

\end{abstract}

\vskip 1 truecm 

\rightline {\em En m\'emoire de  Raymond Stora} 

\section{Introduction}

In quantum field theory, the Schwinger-Dyson equations \cite{1,2} can be understood  \cite{3} as a consequence of the  invariance of the path-integral under field translations. Let $\phi (x) $ denote a set of fields entering the action $S[\phi ]$ and let us introduce the product $X[\phi] = \phi(y_1) \phi(y_2) \cdots \phi(y_n)$.  Invariance of the functional   integration under a field translation    means that the path-integrals over $\phi (x) $ and over $\phi (x) + \eta (x)$ ---where $\eta (x) $ is a given localised classical configuration---  furnish the same result     
\be
\left \langle X[\phi ] \right \rangle = \frac{\int D \phi \; e^{iS[\phi ]} \; X[\phi ] }{\int D \phi \; e^{iS[\phi ]}} = \frac{\int D \phi \; e^{iS[\phi + \eta ]} \; X[\phi + \eta  ] }{\int D \phi \; e^{iS[\phi ]}} \; . 
\label{1.1}
\ee
Therefore, by means of a functional derivative in $\eta (x)$, one obtains 
\bea
 \left \langle \frac {\delta S[\phi ]}{\delta \phi (x)} \, X[\phi] \right \rangle &=& i \left \langle \frac{\delta X[\phi ]} {\delta \phi(x)} \right \rangle \nonumber \\ &=& i \sum_{j=1}^n \, \delta(x-y_j) \left \langle \phi(y_1) \cdots \phi(y_{j-1}) \phi(y_{j+1}) \cdots \phi(y_n) \right \rangle \, . 
\label{1.2}
\eea
Equation (\ref{1.2}) shows that, in the expectation values,  the classical field equations are valid up to the presence of contact terms, which appear on the right-hand-side of equality (\ref{1.2}). 

The Schwinger-Dyson equations can be generalised to the case in which, in the expectation values, one considers ---instead of the field product $X[\phi] = \phi(y_1) \phi(y_2) \cdots \phi(y_n)$---  also composite field operators.  In particular, when $X[\phi] = \delta S[\phi ] / \delta \phi (y_1) \, S[\phi ] / \delta \phi (y_2) \cdots \delta S[\phi ] / \delta \phi (y_n)$, the Schwinger-Dyson equations concern  the expectation values of the products of the composite operator  $\delta S[\phi ] / \delta \phi (x)$ in different points. These expectation values are collectively described by the so-called Schwinger-Dyson functional 
\be
Z_{SD}[b] = \frac{\int D \phi \; e^{iS[\phi ]} \, e^{i \int dx \, b(x) \, \delta S[\phi ] / \delta \phi (x)  }  }{\int D \phi \; e^{iS[\phi ]}} \; ,  
\label{1.3}
\ee
where $b(x)$ denotes a classical source. 

The Schwinger-Dyson equations give significant constraints on the structure of the correlation functions in the quantum Chern-Simons (CS) gauge field theory \cite{4,5}. In the case of the abelian $U(1)$ CS theory, the  Schwinger-Dyson functional for the connected correlation functions of the curvature has the same structure of the action  and  determines  the complete solution \cite{6,7}  of the  theory. In the non-abelian case, the form of the two-point proper vertex is completely specified  \cite{8} by the   Schwinger-Dyson equations.  In this article, the non-abelian CS gauge theory  with gauge group  $SU(N) $ is considered; 
the peculiar form of the Schwinger-Dyson equations is illustrated in a few examples,  and the renormalized Schwinger-Dyson functional $Z_{SD}[\Phi^a_\mu]$ is examined.   
The nonabelian extension of the  result for the abelian theory turns out to be quite  peculiar.   In facts, it is demonstrated that 
$Z_{SD}[\Phi^a_\mu]$ is related with the usual generating functional $Z[J^{a \mu}]$  of the correlation functions of the gauge fields $A^a_\mu$ by some kind of duality transformation.  The exact expression of  the  3-point correlation function of the curvature in the CS theory is derived and its gauge-independence is discussed. 

\section{Renormalized Chern-Simons theory}

The field variation of the  Chern-Simons (CS) action $S [A] $, 
\be
S [A] = {k \over 4 \pi} \int d^3x \, \epsilon^{\mu \nu \tau } \, \Big \{ \mezzo  A^a_\mu \der_\nu A^a_\tau - \sesto \, f^{abc} A^a_\mu A^b_\nu A^c_\tau \Big \} \; , 
\label{2.1}
\ee  
is proportional to the curvature 
\be
\frac {\delta S[A ]}{\delta A^a_\mu (x)} =  \left ( { k \over 8 \pi } \right ) \epsilon^{\mu \nu \tau } F^a_{\nu \tau } = \left ( { k \over 4 \pi } \right ) \epsilon^{\mu \nu \tau }\left ( \der_\nu A_\tau^a - \mezzo \, f^{abc}  A^b_\nu A^c_\tau \right ) \; . 
\label{2.2}
\ee
Therefore the structure of the Schwinger-Dyson equations in the CS theory is originated by 
\be 
\left \langle  \epsilon^{\mu \nu \tau } F^a_{\nu \tau} (x)
\, X[A] \right \rangle =  2 i \left ( { 4 \pi \over k } \right )  \left \langle \frac{\delta X[ A ]} {\delta A^a_\mu (x)} \right \rangle \; . 
\label{2.3}
\ee
In general, equation (\ref{2.3}) must be integrated with the corrections coming from the gauge-fixing procedure. This point will be discussed in a while; for the moment let us proceed with the basic argument.  Since the curvature $F^a_{\mu \nu}$ transforms covariantly under a local gauge transformation, the contact terms in the Schwinger-Dyson equations can combine to produce a gauge-invariant expression.   In particular,  according to the rule shown in equation ({\ref{2.3}),  the 3-points function of the curvature  should be given by  the gauge-invariant combination 
\be
 \left \langle  \epsilon^{\mu \alpha \beta } F^a_{\alpha \beta } (x)  \, \epsilon^{\nu \gamma  \delta } F^b_{\gamma \delta } (y) \, \epsilon^{\tau \sigma \xi } F^c_{\sigma \xi } (z) \right \rangle = 16 \left ( {4 \pi \over k} \right )^2  f^{abc} \, \epsilon^{\mu \nu \tau}
\, \delta^3 (x-y) \, \delta^3 (z - y) \; . 
\label{2.4}
\ee
In the CS theory, the  value of the 3-points correlation function of the curvature corresponds to the gauge-invariant expression (\ref{2.4}), which is proportional to the structure-constants tensor $f^{abc}$ of the gauge group divided by the  square of the coupling constant $k$. As it will be shown below, the expression appearing in equation (\ref{2.4}) should have a  gauge-independent meaning because it can also be obtained in the limit of vanishing gauge-fixing.    As a check, one can easily verify the validity of equation (\ref{2.4}) to lowest  orders of perturbation theory when the CS theory is formulated in ${\mathbb R}^3$.    
 
In order to proceed with the derivation of the properties of the renormalized Schwinger-Dyson functional in the CS theory, one needs to specify the gauge-fixing procedure and the renormalization conditions. Let us consider the perturbative approach to the CS theory  in ${\mathbb R}^3$. Really, in the following argument ${\mathbb R}^3$ can be replaced by a generic 3-manifold $M$ which is a homology sphere because, in this case, the field variables have no zero modes \cite{6} and the standard perturbative expansion is well defined. In the Landau gauge,  the gauge-fixing term \cite{9} is given by    
 \be
S_{\phi \pi} = {k \over 4 \pi} \int d^3x \Big \{ - B^a\der^\mu A^a_\mu + \der^\mu {\overline c}^a   \left ( \der_\mu c^a - f^{abc} A^b_\mu c^c \right ) \Big \} \; . 
\label{2.5}
\ee
Let $\Gamma$ be the renormalized effective action of the CS theory;  $i \, \Gamma$ is given by  the sum of the one-particle-irreducible diagrams with external legs represented   by classical fields.  

$\Gamma $ can be  computed by means of various techniques; one convenient method is the standard quantum field theory  procedure which is called the {\it renormalized perturbation theory} in the Peskin-Schroeder book \cite{10}.  Of course, any other renormalization method leads to the same physical conclusions;    the use of  renormalized perturbation theory is quite instructive because the basic concepts of the renormalization clearly emerge. In renormalized perturbation theory, the values of all the parameters entering the lagrangian coincide with the renormalized values, and the so-called local counterterms cancel  precisely all the possible contributions  to these parameters which are found in the loop expansion. In this way, the  normalization conditions \cite{3,10} are indeed satisfied  to all orders of perturbation theory, as it must be.  

 Renormalized perturbation theory represents one of the fundamental constituents of the theory of quantized fields  \cite{11,12}; this subject was of particular interest for Raymond Stora   \cite{13}. So I will elaborate a bit on this issue in the context of the quantum CS field theory. At the beginning of the years 90's, with Raymond we had fruitful discussions on this matter. 
   
The renormalization process of the CS action (\ref{2.1}) concerns two parameters:  the wave function normalization  and the  coupling constant. Actually, as in any gauge theory,  because of the gauge invariance one of the normalization conditions  is superfluous \cite{3}; therefore, in our case,  only  one parameter needs to be specified. The CS normalization conditions can then be expressed as  

\begin{enumerate}[(i)]
\item  $\Gamma $ is invariant under BRST transformations \cite{14} which act on the fields  ---appearing in the action--- according to   
\bea
&& \delta \,  A^a_\mu (x) = \der_\mu c^a (x) -  f^{abd} A_\mu^b(x) c^d(x) \quad , \quad \delta \, c^a (x) = \mezzo  f^{abd} c^b (x) c^d (x)  \nonumber  \\ && \delta \, {\overline c}^a(x) = - B^a (x) \quad , \quad  \delta \, B^a(x) = 0 \; .
\label{2.6}
\eea

\item The renormalized coupling constant $k$ is specified by the 2-point proper vertex at vanishing momenta. More precisely, let $\Gamma_{AA}$ be  the term of  the expansion of $\Gamma $ in powers of the fields  which is quadratic in the field $A^a_\mu$, 
\be
\Gamma_{AA} = \mezzo \int {d^3 p \over (2 \pi )^3}  \, \widetilde A^a_\mu (p) \widetilde A^b_\nu (-p)   \, {\Gamma^{(2)}}^{\mu \nu}_{a b } (p) \; . 
\label{2.7}
\ee
Then the normalization condition takes the form 
\be
\lim_{p \rightarrow 0} {\Gamma^{(2)}}^{\mu \nu}_{a b } (p) =  \left ( {k \over 4 \pi } \right )  \delta_{ab} \, \epsilon^{\mu \nu \tau} (-i p_\tau  ) \; . 
\label{2.8}
\ee
\end{enumerate}

Since there are no gauge anomalies in three dimensions,  condition (i) is well-suited, and condition (ii) gives the definition of the coupling constant. Condition (\ref{2.8}) is the analogue of the normalization condition in quantum electrodynamics or in Yang-Mills theory \cite{3,10} and  admits the following  equivalent formulation. Let $\Gamma_{AAA}$ be  the term of  the expansion of $\Gamma $ in powers of the fields  which is cubic in the field $A^a_\mu$, 
\be
\Gamma_{AAA} = \sesto \int {d^3 p_1 \over (2 \pi )^3}  {d^3 p_2 \over (2 \pi )^3} \, \widetilde A^a_\mu (p_1)  \widetilde A^b_\nu (p_2)  \widetilde A^c_\tau (-p_1 - p_2)   \, {\Gamma^{(3)}}^{\mu \nu \tau}_{a b c} (p_1 , p_2) \; . 
\label{2.9}
\ee
Because of the BRST invariance, condition (\ref{2.8}) is equivalent to 
\be
\lim_{\substack{p_1 \rightarrow 0 \\ 
p_2 \rightarrow 0}}
{\Gamma^{(3)}}^{\mu \nu \tau}_{a b c} (p_1 , p_2) =  - \left ( {k \over 4 \pi } \right )  f_{abc} \, \epsilon^{\mu \nu \tau}  \; . 
\label{2.10}
\ee
In renormalized perturbation theory,  the real parameter $k$ multiplying the action (\ref{2.1}) denotes the renormalized CS coupling constant  that takes integer values ($k = 1,2,3,...$) and receives no corrections.  Consequently,  the coupling constant entering equations (\ref{2.3}) and  (\ref{2.4}) represents the renormalized coupling constant. All the equations of the present article are expressed in terms of the renormalized coupling constant $k$. 

The exact scale invariance \cite{15,16} of $\Gamma $ implies that, in the correlation functions, typical logarithms of the momenta cannot appear; the CS theory is actually finite because, at the regularized level, all the potential divergences cancel.  The  BRST symmetry together with a vector supersymmetry  invariance \cite{17,18} of the action  specify the value of  $\Gamma_{AA}$ uniquely; in facts  the   Sorella-Piguet non-renormalization theorem  \cite{19} states     

\newtheorem{theorem}{Theorem}
\begin{theorem}[Sorella--Piguet]
{\it In the CS theory with Landau gauge fixing, one has}
\be
\Gamma_{AA} = {k \over 4 \pi} \int d^3x \, \mezzo \, \epsilon^{\mu \nu \tau } \, A^a_\mu (x) \der_\nu A^a_\tau (x) \; . 
\label{2.11}
\ee
\end{theorem}

\noindent{\bf {Proof}.} In Landau gauge, the total action $S_{TOT} = S + S_{\phi \pi}$ is invariant under a vector supersymmetry [17,18]. 
If, at a given order of perturbation theory, this  vector supersymmetry  has no anomalies, one can define a renormalized effective action such that the Ward identities coming from this  supersymmetry are satisfied.   
Then  the Schwinger-Dyson equations combined with the Ward identities and BRST  invariance  imply  \cite{8,19}  that  the  2-point proper vertex has the form $\Gamma_{AA} \propto \int A^a \wedge dA^a$.  All the explicit computations at one loop and at two loops show that the vector supersymmetry in the CS theory has no anomalies. Actually, Sorella and Piguet have demonstrated \cite{19} that this vector supersymmetry is not anomalous to all orders of perturbation theory. Therefore the {\em exact}  2-point proper vertex  is given by $\Gamma_{AA} = \beta \int A^a \wedge dA^a$ where $\beta $ is a real parameter.   The value of this parameter is specified precisely by the normalization condition (ii), and so $\beta = k / 8 \pi $. {\hfill \cvd}

\medskip

 A consequence of the non-renormalization  theorem is that the Feynman propagator  for the components of $A^a_\mu $ actually coincides with the dressed propagator,  
 \be
\Delta^{ab}_{\mu \nu }(x -y) = \WT{A^a_\mu(x)\> A^b}{\! \null_\nu}(y) =  \delta^{a b} \left ( { 4 \pi \over  k } \right )  \int {d^3 p \over (2 \pi )^3} \,  {e^{ip \cdot x }\, \varepsilon_{\nu \mu \lambda } p^\lambda \over p^2}  \; . 
\label{2.12}
\ee
Similarly, the ghosts propagator also gets no loop corrections \cite{8, 19} as a consequence of the vector supersymetry 
\be
\Delta^{ab}(x-y) = \WT{c^a(x)\, {\overline c}^b}(y) = i \delta^{ab}  \left ( { 4 \pi \over  k } \right ) \int {d^3 p \over (2 \pi )^3} {e^{ip \cdot x} \over p^2}  \; . 
\label{2.13} 
\ee
Equation (\ref{2.11}) is in agreement with the outcome of  renormalized perturbation theory. The result (\ref{2.11}) is also consistent with the topological character of the CS theory because the Gauss linking number does not admit nontrivial radiative corrections. 
 
\subsection{Normalization conditions} 
Let us concentrate on renormalizable  quantum field theories.  In each model, the normalization conditions specify the finite (renormalized) values of all the parameters of the model which enter the correlation functions,  the transition amplitudes, and the observables  in general. For instance, in quantum electrodynamics (QED) the cross sections for the electromagnetic scatterings between electrons and photons depend on the electromagnetic renormalized  coupling constant  $\alpha_{em}$. The finite value of  $\alpha_{em} $ which is measured in laboratories ($\alpha_{em} \simeq 1 /137$) can be specified  by a suitable normalization condition \cite{3,10} that must be satisfied by the 2-point proper vertex of the electromagnetic vector potential $A_\mu$. Really one needs to specify  a complete set of normalization conditions but, in order  to simplify the exposition,  let us concentrate on $\alpha_{em}$.  Since the normalization conditions give the definition of the coupling constant, the normalization conditions must be satisfied at each order of perturbation theory.   
 Thus the value of $\alpha_{em} $ does not receive loop corrections; there is no shift in $\alpha_{em} $.  In facts, the value of $\alpha_{em}$ is one of the  inputs of the theory, whereas  the predictions are obtained by computing  how the other observables depend on $\alpha_{em}$.  This is why all  the textbooks on quantum field theory do not contain  computations of the radiative corrections to $\alpha_{em} $;   they contain instead computations showing how the transition amplitudes, cross sections, etc.  depend on $\alpha_{em}\simeq 1/137$. 

In order to construct the renormalized effective action $\Gamma $, certain renormalization  methods introduce, in the intermediate steps of the regularization/renormalization  procedure, one or several regulator cut-offs and bare lagrangian parameters. So one could imagine that the dependence of the renormalized coupling constant on the bare coupling constant is a meaningful issue. But this is not the case, because the bare coupling constant is not  observable;  this point is precisely the fundamental discovery in renormalization theory.  It turns out that the dependence of the renormalized coupling constant on the bare coupling constant is not unique  and, in the perturbative expansion, it  can be  modified \cite{3,10,11,12}   without any observable or physical consequence. 
Instead, how the observables of each field theory model  depend on the renormalized coupling constant ---that satisfies the normalization conditions--- is uniquely determined. For this reason, renormalized perturbation theory is very instructive, because in this procedure there are no imagined bare parameters at all.   
   
   Similarly, in the CS theory one needs to specify the value of the renormalized coupling constant $k$; this can  be done  by means of the condition (\ref{2.8}).  Gauge invariance requires that the renormalized value of $k$ must be an  integer, $k = 1,2,3,.... $;  precisely like $\alpha_{em}$ in QED,  this integer $k$ does not get   loop corrections  \cite{9,20}  because the normalization conditions must be satisfied at each order of perturbation theory.   Thus there is no shift of  the coupling constant $k$ in the CS theory simply because there is no shift of $\alpha_{em}$ in QED, and there is no shift  of the renormalized coupling constant in any renormalizable quantum field theory.  
   
What  is remarkable in the CS theory is that, according to the non-renormalization theorem ---and  in agreement with the two-loops computations \cite{9}---,   the {\em entire} 2-point proper vertex  is not modified by radiative corrections. Consequently, the CS  vacuum polarization is vanishing to all orders of perturbation theory; this implies that, in the CS theory, the ordinary Schwinger-Dyson equations gets somewhat modified and assume a simplified form, as it will be illustrated in the next section. 
  
 \section{Equations for proper vertices}
 
 In the gauge-fixed CS theory, the  Schwinger-Dyson equations for the proper vertices ---containing at least one gauge field---  can be obtained \cite{3} from the equation 
 \be
  \left \langle \left [ \frac {\delta S_{TOT}}{\delta A^a_\mu (x)}  + J^{a \mu} (x) \right ] \, e^{i \int J\cdot \phi }\right \rangle  =0 \; , 
  \label{3.1}
 \ee
which is a particular case of equation (\ref{1.2}). The expectation value (\ref{3.1})  must be  computed by using  the gauge-fixed CS action 
\be 
S_{TOT } = S + S_{\phi \pi} \; . 
\label{3.2}
\ee
Moreover, 
 \be
 e^{i \int J\cdot \phi }= \exp \left \{ i \! \int d^3x \left [   J^{a \mu }(x) A^a_\mu (x) + L^a (x)B^a (x)+ \eta^a (x){\overline c}^{\, a }(x) + {\overline \eta}^{\, a} (x) c^a (x)\right ] \right \} \; , 
 \label{3.3}
 \ee
 in which the classical sources of commuting type $J^{a \mu}$, $L^a$  and of anticommuting type $\eta^a$, ${\overline \eta}^{\, a}$  have been introduced. By means of the generating functional $W$ of the connected correlation functions, 
 \be
e^{i W [J^{a \mu}, L^a, \eta^a , {\overline \eta}^{\, a}] }= \left \langle e^{i \int J\cdot \phi }
\right \rangle \; ,  
 \label{3.4}
\ee
equation (\ref{3.1}) can be rewritten as 
 \bea
 0 &=& \epsilon^{\mu \nu \tau } \der_\nu {\delta W\over \delta J^{a \mu }(x)}  - \mezzo \epsilon^{\mu \nu \tau } f^{abc}  {\delta W\over \delta J^{b \nu }(x)}{\delta W\over \delta J^{c \tau }(x)} + \imezzi \epsilon^{\mu \nu \tau }  f^{abc} {\delta^2 W \over \delta J^{b \nu }(x) J^{c \tau }(x)}\nonumber \\
&& + f^{abc} \, \der^\mu_x {\delta W \over \delta \eta^b(x)}{\delta W \over \delta {\overline \eta}^c (x) } -i f^{abc} \, \der^\mu_x {\delta^2 W \over \delta \eta^b(x){\overline \eta}^c(y)}\Bigg |_{x=y}\nonumber \\
&& + \der^\mu {\delta W \over \delta L^a(x)} + \left ( {4 \pi \over k}\right ) J^{a \mu }(x)  \; . 
\label{3.5}
 \eea
Now one can  put
\be
W = \Gamma + \int  \left [   J^{a \mu }A^a_\mu  + L^a B^a+ \eta^a {\overline c}^{\, a } + {\overline \eta}^{\, a}  c^a \right ]  
\label{3.6}
\ee
in equation (\ref{3.5}), where the Legendre transform $\Gamma $ of $W$ satisfies 
\bea
{\delta \Gamma \over \delta A^a_\mu (x)} = - J^{a \mu }(x) \quad &,& \quad {\delta \Gamma \over \delta B^a_\mu (x)} = - L^{a }(x) \; , \nonumber \\
{\delta \Gamma \over \delta {\overline c}^a (x)} = \eta^{a }(x) \quad &,& \quad {\delta \Gamma \over \delta  c^a (x)} = - {\overline \eta}^{a }(x) \; . 
\label{3.7}
\eea
The various functional derivatives of the expressions appearing in equation (\ref{3.5}) give rise to a sequence of relations for the proper vertices and the dressed propagators.  These relations assume a quite peculiar form because of the validity of the non-renormalization theorem.   As depicted in Figure~1, equation (\ref{2.11}) states that the  vacuum polarization for the gauge fields  is vanishing. 
  
\vskip 0.7 truecm

\centerline {
\begin{tikzpicture} [scale=0.6] [>=latex]
\draw [very thick , pattern=north east lines ] (0,0) circle (0.8); 
\draw [very thick]  (-2,0) -- (-0.8,0); 
\draw [very thick]  (0.8,0) -- (2,0);
\draw [very thick] (4.2,0) -- (8.2,0); 
\draw [very thick] (5.9,0.3) -- (6.5,-0.3); 
\draw [very thick] (5.9,-0.3) -- (6.5,0.3); 
\node at (3.2,0) {$  =  $};
\end{tikzpicture}
}

\vskip 0.3 truecm
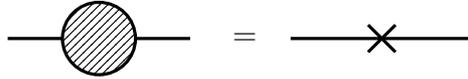
\centerline {{Figure 1.} {Vanishing of the vacuum polarization for the connection.}}

\vskip 0.7 truecm

\noindent Because of the CS vector supersymmetry, also the 2-point proper vertex for the ghosts fields receives no corrections; in agreement with equation (\ref{2.13}) one has then 
\be
\Gamma_{{\overline c} c} = {k \over 4 \pi} \int d^3x \, \der^\mu {\overline c}^a (x) \, \der_\mu c^a (x)  \; . 
\label{3.8}
\ee
Equation (\ref{3.8}) can be  represented by the diagram shown in Figure~2. 

\vskip 0.7 truecm

\centerline {
\begin{tikzpicture} [scale=0.6] [>=latex]
\draw [very thick , pattern=north east lines ] (0,0) circle (0.8); 
\draw [very thick , densely dashed ]  (-2,0) -- (-0.8,0); 
\draw [very thick , densely dashed ]  (0.8,0) -- (2,0);
\draw [very thick , densely dashed ] (4.2,0) -- (8.2,0); 
\draw [very thick] (5.9,0.3) -- (6.5,-0.3); 
\draw [very thick] (5.9,-0.3) -- (6.5,0.3); 
\node at (3.2,0) {$  =  $};
\end{tikzpicture}
}

\vskip 0.3 truecm
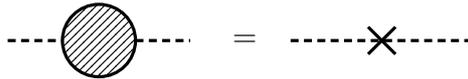
\centerline {{Figure 2.} {Vanishing of the vacuum polarization for the ghosts.}}

\vskip 0.7 truecm

\noindent By taking one functional  derivative with respect to  $A^d_\nu (y)$ of the functions appearing in equation (\ref{3.5}), and by letting all the sources vanish, one obtains the relation 
\bea
&&\int d^3u \, d^3 w {\delta^3 \Gamma \over \delta A^d_\nu (y) \delta A^e_\alpha (u) \delta A^h_\beta (w)} \Delta^{e b}_{\alpha \tau} (u - x) \Delta^{h c}_{\beta \lambda } (w - x) 
\epsilon^{\mu \tau \lambda } f^{abc} = \nonumber \\ 
&& \qquad = 2 \int d^3u \, d^3 w {\delta^3 \Gamma \over \delta A^d_\nu (y) \delta {\overline c}^{\, e} (u) \delta c^h (w)}\, \der^\mu_x \Delta^{e b} (u - x) \Delta^{h c} (w - x) 
 f^{abc}\; , 
 \label{3.9}
\eea
which can be represented as shown in Figure~3. 
  
\vskip 0.7 truecm

\centerline {
\begin{tikzpicture} [scale=0.5] [>=latex]
\draw [very thick , pattern=north east lines ] (-0.2,0) circle (0.8); 
\draw [very thick]  (-2,0) -- (-1,0); 
\draw [very thick] (1.4,0)  ++(-150: 1.3) arc (-150:150:1.3);
\draw [very thick]  (2.7,0) -- (3.7,0); 
\draw [very thick , pattern=north east lines ] (8.4,0) circle (0.8); 
\draw [very thick]  (6.6,0) -- (7.6,0); 
\draw [very thick , densely dashed ] (10,0)  ++(-150: 1.3) arc (-150:150:1.3);
\draw [very thick]  (11.3,0) -- (12.3,0); 
\node at (5.2,0) {$  =  \; \; - $};
\end{tikzpicture}
}

\vskip 0.3 truecm
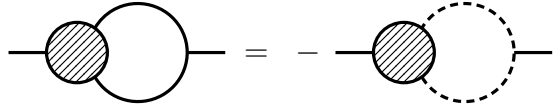
\centerline {{Figure 3.} {Schwinger-Dyson equation for the 2-point vertex of the gauge fields.}}

\vskip 0.7 truecm

\noindent In deriving equation (\ref{3.9}), the relations illustrated in Figure~1 and Figure~2 have been taken into account. A further derivative in the gauge field of the functionals  entering equality (\ref{3.5}) leads to the relation shown in Figure~4. 

\vskip 0.7 truecm

\centerline {
\begin{tikzpicture} [scale=0.7] [>=latex]
\draw [very thick , pattern=north east lines ] (0,0) circle (0.6); 
\draw [very thick]  (-1,-1) -- (-0.4,-0.4); 
\draw [very thick]  (1,-1) -- (0.4,-0.4); 
\draw [very thick]  (0,0.6) -- (0,1.3);
\node at (2,0) {$  =  $};
\draw [very thick]  (4,0) -- (4,1);
\draw [very thick]  (3,-1) -- (4,0); 
\draw [very thick]  (5,-1) -- (4,0); 
\node at (6,0) {$  +  $};
\draw [very thick , pattern=north east lines ] (8,-1) circle (0.6); 
\draw [very thick]  (7,-2) -- (7.6,-1.4); 
\draw [very thick]  (9,-2) -- (8.4,-1.4); 
\draw [very thick] (8,0)  ++(-57:1) arc (-57:235:1);
\draw [very thick]  (8,1) -- (8,1.8); 
\node at (10,0) {$  +  $};
\node at (10.65,0) {$  2 $};
\draw [very thick , pattern=north east lines ] (11.6,-1) circle (0.6); 
\draw [very thick , pattern=north east lines ] (13.6,-1) circle (0.6); 
\draw [very thick]  (10.56,-2) -- (11.16,-1.4); 
\draw [very thick]  (14.65,-2) -- (14.05,-1.4); 
\draw [very thick] (12.6,0)  ++(-22:1.1) arc (-22:202:1.1);
\draw [very thick] (12.6,0)  ++(247:1.1) arc (247:292:1.1);
\draw [very thick]  (12.6,1.1) -- (12.6,1.9); 
\node at (6,-5) {$  +  $};
\draw [very thick , pattern=north east lines ] (8,-6) circle (0.6); 
\draw [very thick]  (7,-7) -- (7.6,-6.4); 
\draw [very thick]  (9,-7) -- (8.4,-6.4); 
\draw [very thick, densely dashed ] (8,-5)  ++(-57:1) arc (-57:235:1);
\draw [very thick]  (8,-4) -- (8,-3.2); 
\node at (10,-5) {$  +  $};
\node at (10.65,-5) {$   2 $};
\draw [very thick , pattern=north east lines ] (11.6,-6) circle (0.6); 
\draw [very thick , pattern=north east lines ] (13.6,-6) circle (0.6); 
\draw [very thick]  (10.56,-7) -- (11.16,-6.4); 
\draw [very thick]  (14.65,-7) -- (14.05,-6.4); 
\draw [very thick, densely dashed ] (12.6,-5)  ++(-22:1.1) arc (-22:202:1.1);
\draw [very thick, densely dashed ] (12.6,-5)  ++(247:1.1) arc (247:292:1.1);
\draw [very thick]  (12.6,-3.93) -- (12.6,-3.1); 
\end{tikzpicture}
}

\vskip 0.5 truecm

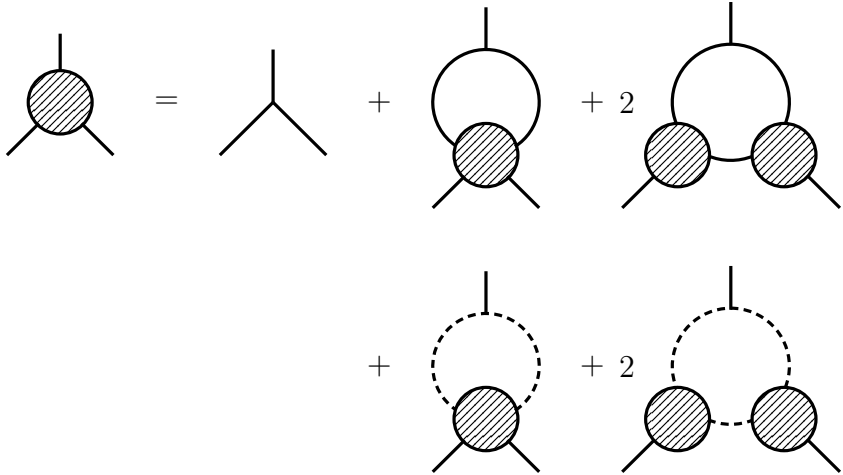
\centerline {{Figure 4.} {Schwinger-Dyson equation for the 3-point vertex of the gauge fields.}}

\vskip 0.7 truecm

\noindent It is tempting to conjecture that, because of their  simplified structure, the Schwinger-Dyson equations ---for the CS theory in ${\mathbb R}^3$--- can be solved.  So far, only the first few terms of the perturbative expansion have been explored.

\section{Schwinger-Dyson functional}

Another basic feature of the CS theory is that 
the interaction lagrangian ${\cal L}_I$ is a cubic function of the gauge fields, {\it i.e.}   ${\cal L}_I \sim A^3$.  Consequently, the   Schwinger-Dyson functional is strictly related with the generating functionals of the correlation functions, as it will be shown in this section. 
The Schwinger-Dyson functional $Z_{SD}[\Phi^a_\mu ]$ for the gauge-fixed CS theory in ${\mathbb R}^3$ is defined by  
\be
 Z_{SD}[\Phi^a_\mu ] =   \left \langle \exp \left ( i \int d^3x \, \Phi^a_\mu(x)  \left [ \delta (S + S_{\phi \pi}) / \delta A_\mu^a(x) \right ] \right ) \right \rangle \; , 
 \label{4.1}
 \ee
where $ \Phi^a_\mu(x)$ is a classical source field. Let $Z[J^{a \mu} ]$ be the standard generating functional  of the renormalized correlation functions of the field $A_\mu^a $, 
\be
Z[J^{a \mu} ] = \left \langle \exp \left ( i \int d^3x \, J^{ a \mu } (x)  A_\mu^a(x)  \right ) \right \rangle \; . 
 \label{4.2}
\ee
 
 \newtheorem{prop}{Proposition}
\begin{prop}
{\it The functionals $Z_{SD}[\Phi^a_\mu ]$ and $Z[J^{a \mu} ]$ are related by a duality transformation according to}
\be
Z_{SD}[\Phi^a_\mu ] = e^{iR[\Phi^a_\mu]} \, Z[\widetilde J[\Phi]^{a \mu}  ]\; , 
\label{4.3}
\ee
where
\be
R[\Phi^a_\mu] = - {k \over 4 \pi} \int d^3x \, \epsilon^{\mu \nu \tau } \left \{ \mezzo \Phi^a_\mu \der_\nu \Phi^a_\tau + \terzo f^{abc} \Phi^a_\mu \Phi^b_\nu \Phi^c_\tau \right \}\; , 
\label{4.4}
\ee
and
\be
\widetilde J[\Phi]^{a \mu}(x) = {k \over 8 \pi } \epsilon^{\mu \nu \tau} \, f^{abc} \Phi_\nu^b(x) \Phi^c_\tau (x)\; . 
\label{4.5}
\ee

\end{prop}
 
 \noindent{\bf {Proof}.} Relation (\ref{4.3}) is a consequence of the invariance of the path-integral under a field translation, as indicated in equation (\ref{1.1}). Indeed one has 
 \bea
 S_{TOT} [A + \Phi] &=& S_{TOT} [A] + \int d^3x \, \Phi^a_\mu(x)  \left [ \delta S_{TOT} / \delta A_\mu^a(x) \right ] \nonumber \\
 && + S[\Phi] - {k \over 8 \pi }  \int d^3x \,  \epsilon^{\mu \nu \tau} f^{abc}   A_\mu^a (x) \Phi_\nu^b (x) \Phi_\tau^c(x) \; . 
 \label{4.6}
 \eea
This equation can be written in the form 
\bea
S_{TOT} [A] + &&{\hskip -0.7 truecm} \int d^3x \, \Phi^a_\mu(x)  \left [ \delta S_{TOT} / \delta A_\mu^a(x) \right ] =  R[\Phi^a_\mu] +  S_{TOT}[A+\Phi] \nonumber \\ 
&& + {k \over 8 \pi } \int d^3x \,  \epsilon^{\mu \nu \tau} f^{abc}   \left ( A_\mu^a (x) + \Phi^a_\mu (x) \right ) \Phi_\nu^b (x) \Phi_\tau^c(x)  \; . 
\label{4.7}
\eea
 Consequently, the integration over the field variables gives 
 \bea
 \frac{\int D (\hbox{fields}) \; e^{iS_{TOT}[A ] + i\int \Phi^a_\mu (\delta S_{TOT} / \delta A^a_\mu )}}{\int D (\hbox{fields}) \; e^{iS_{TOT}}} &=&  e^{i R[\Phi^a_\mu ] } \times  \nonumber \\ 
 && {\hskip - 2.5 truecm} \times \frac{\int D (\hbox{fields}) \; e^{iS_{TOT}[A + \Phi] + i\int (A^a_\mu+\Phi^a_\mu) \widetilde J^{a \mu} }}{\int D (\hbox{fields}) \; e^{iS_{TOT}}} \; ,  
 \label{4.8}
 \eea
 where $D (\hbox{fields}) = DA_\mu^a \, DB^c \,  D{\overline c}^b \, D c^d$. 
 Therefore one obtains 
 \be 
 \left \langle \exp \left ( i \int d^3x \, \Phi^a_\mu  (x) \left [ \delta S_{TOT} / \delta A_\mu^a (x)\right ] \right ) \right \rangle = e^{i R[\Phi^a_\mu ] } 
  \left \langle \exp \left ( i \int d^3x \, \widetilde J[\Phi ]^{ a \mu } (x) A_\mu^a  (x)\right ) \right \rangle \; ,  
  \label{4.9}
 \ee
and this concludes the proof. {\hfill \cvd}

\medskip
 
Note that the validity of equation (\ref{4.3}) is not restricted to the case of the Landau gauge. Provided that $S_{\phi \pi}$ is a linear function of $A_\mu^a$, the result  (\ref{4.3}) follows where, independently of the particular choice  of the gauge-fixing, $R[\Phi^a_\mu ] $ takes the form shown in equation (\ref{4.4}). 
In the $f^{abc} \rightarrow 0$ limit, one recovers the results of the abelian CS theory; indeed in this limit one finds $\widetilde J[\Phi]^{a \mu}(x) \rightarrow 0$ and the  term of 
$R[\Phi^a_\mu ] $ which is quadratic in the source field $\Phi^a_\mu (x)$ determines the Schwinger-Dyson functional of the $U(1)$ CS theory \cite{6}. 

The composite operator $\delta S_{TOT} / \delta A_\mu^a (x)$ has dimension 2 and enters the construction of the renormalized CS effective action; so for smooth classical source $\Phi^a_\mu (x)$, equation  (\ref{4.3}) is expected to give  the relation between the renormalized Schwinger-Dyson functional  $Z_{SD}[\Phi^a_\mu ] $ and the renormalized generating functional $Z[J^{a \mu}]$. 
 
In order to recover equation (\ref{2.4}), let us consider the case in which the action $S_\lambda $ of the gauge-fixed theory is given by 
\be
S_\lambda = S + \lambda S_{\phi \pi} \; , 
\label{4.10}
\ee 
where $S$ and $S_{\phi \pi}$ are shown in equations (\ref{2.1}) and (\ref{2.5}), and $\lambda $ is a real parameter. Since $S$ and $S_{\phi \pi}$ are separately BRST-invariant, for any fixed  value of $\lambda $ the action $S_\lambda$ is invariant under the BRST transformations (\ref{2.6}). Let $\langle Y \rangle_\lambda  $ denote the normalised expectation value of a generic field function $Y$ in the particular field theory model that is specified  by the action $S_\lambda$. Since the gauge fixing lagrangian  is a linear function of the gauge field $A^a_\mu$,  one obtains 
 \be 
 \left \langle \exp \left ( i \int d^3x \, \Phi^a_\mu   \left [ \delta S_\lambda / \delta A_\mu^a \right ] \right ) \right \rangle_\lambda = e^{i R[\Phi^a_\mu ] } 
  \left \langle \exp \left ( i \int d^3x \, \widetilde J[\Phi ]^{ a \mu }  A_\mu^a  \right ) \right \rangle_\lambda \; .   
  \label{4.11}
 \ee
Let us now consider the expansion in powers of  $\Phi^a_\mu$ of  the functions which appear  in equation (\ref{4.11}); we are interested in the coefficient of the term which is cubic in $\Phi^a_\mu$. Because
\be
{\delta S_\lambda \over  \delta A_\mu^a(x)} =  \left ( {k \over 8 \pi} \right )  \left ( \epsilon^{\mu \alpha \beta} F^a_{\alpha \beta} (x) +  2 \lambda \der^\mu B^a + 2\lambda f^{apq} \der^\mu {\overline c}^{\, p }(x) c^q(x) \right ) \; , 
\label{4.12}
\ee
the left-hand-side of equation (\ref{4.11})  gives 
\bea
 \hbox{ left } \big |_\lambda  &=&  \left ( {ik \over 8 \pi} \right )^3  
  \Big \langle   \left [ \epsilon^{\mu \alpha \beta} F^a_{\alpha \beta} (x) +  2 \lambda \der^\mu B^a (x) + 2\lambda f^{apq} \der^\mu {\overline c}^{ p }(x) c^q(x)  \right ] \times \nonumber \\ 
  &&{\hskip  0.8 truecm}
\times  \left [ \epsilon^{\nu \gamma \delta} F^b_{\gamma \delta} (y) +  2 \lambda \der^\nu B^b (y)+ 2\lambda f^{brs} \der^\nu {\overline c}^{ r }(y) c^s(y)  \right ] \times \nonumber \\
&&{\hskip  0.8 truecm}
\times   \left [ \epsilon^{\tau \sigma \xi} F^c_{\sigma \xi} (z) +  2 \lambda \der^\tau B^c (z) + 2\lambda f^{c tv} \der^\tau {\overline c}^{ t }(z) c^v(z)  \right ] \Big \rangle_\lambda \; . 
  \label{4.13}
 \eea
Since the nontrivial part of $\langle e^{i \int J^{a \mu} A^a_\mu }\rangle $ is at least quadratic in $J^{a \mu}$, only $R[\Phi^a_\mu ] $ contributes to the cubic term in $\Phi^a_\mu $ on the right-hand-side of equation (\ref{4.11}), 
\be
 \hbox{ right } \big |_\lambda 
 =  - 2 i \left ( {k \over 4 \pi} \right ) f^{abc} \epsilon^{\mu \nu \tau} \delta^3 (x-y) \delta^3 (z - y )\; . 
 \label{4.14}
 \ee
 Let us now consider the $\lambda \rightarrow 0 $ limit. Expression (\ref{4.14}) does not depend on $\lambda $. Whereas, 
 \be
 \lim_{\lambda \rightarrow 0}  \hbox{ left } |_\lambda = -  {i\over 8}  \left ( {k \over 4 \pi} \right )^3 
  \left \langle  \epsilon^{\mu \alpha \beta } F^a_{\alpha \beta } (x)  \, \epsilon^{\nu \gamma  \delta } F^b_{\gamma \delta } (y) \, \epsilon^{\tau \sigma \xi } F^c_{\sigma \xi } (z) \right \rangle
  \label{4.15}
 \ee
  denotes the 3-points correlation function of the curvature in the limit of vanishing gauge-fixing.  
 In the $\lambda \rightarrow 0 $ limit, equality (\ref{4.11}) then implies 
  \be
  f^{abc} \epsilon^{\mu \nu \tau} \delta^3 (x-y) \delta^3 (z - y ) = {1\over 16} \left ( {k \over 4 \pi} \right )^2 
  \left \langle  \epsilon^{\mu \alpha \beta } F^a_{\alpha \beta } (x)  \, \epsilon^{\nu \gamma  \delta } F^b_{\gamma \delta } (y) \, \epsilon^{\tau \sigma \xi } F^c_{\sigma \xi } (z) \right \rangle \; , 
\label{4.16}
\ee
 which coincides with equation (\ref{2.4}).  
  Thus, similarly to the expectation values of the gauge-invariant observables in any gauge theory,  equation (\ref{2.4}) can formally be obtained in the limit in which the gauge-fixing lagrangian term is absent.

 \vskip 1.4 truecm 

\bibliographystyle{amsalpha}

\end{document}